\documentclass[prc,twocolumn,showpacs,preprintnumbers,amsmath,amssymb]{revtex4}
\usepackage{txfonts}
\usepackage{amssymb}
\usepackage{graphicx}

\begin{document}

\title{Sub-barrier fusion of $^{32}$S+$^{90,96}$Zr: semi-classical
coupled-channels approach}
\author{H.~Q.~Zhang$^{1}$}
\author{C.~J.~Lin$^{1}$}
\author{F.~Yang$^{1}$}
\author{H.~M.~Jia$^{1}$}
\author{X.~X.~Xu$^{1}$}
\author{F.~Jia$^{1}$}
\author{Z.~D.~Wu$^{1}$}
\author{S.~T.~Zhang$^{1}$}
\author{Z.~H.~Liu$^{1}$}
\author{A. Richard$^{1,2,3}$}
\author{C. Beck$^{2}$}

\affiliation{ $^{1}$China Institute of Atomic Energy, P.~O.~Box
275(10), Beijing 102413, P.~R.~China\\
$^{2}$Institut Pluridisciplinaire Hubert Curien, UMR 7178,
CNRS-IN2P3, and Universit\'{e} de Strasbourg (UdS), F-67037 Strasbourg
Cedex 2, France\\
$^{3}$Ecole Nationale Sup\'{e}rieure de Physique de Strasbourg
(ENSPS), F-67400 Illkirch, France\\}

\date{\today}

\begin{abstract}
The fusion excitation functions have been measured with rather good accuracy
for $^{32}$S+$^{90}$Zr and $^{32}$S+$^{96}$Zr near and below the Coulomb
barrier. The sub-barrier cross sections for $^{32}$S+$^{96}$Zr are much
larger compared with $^{32}$S+$^{90}$Zr. Semi-classical coupled-channels
calculations including two-phonon excitations are capable to describe
sub-barrier enhancement only for $^{32}$S+$^{90}$Zr. The remaining
disagreement for $^{32}$S+$^{96}$Zr comes from the positive \emph{Q}-value
intermediate neutron transfers in this system. The comparison with
$^{40}$Ca+$^{96}$Zr suggests that couplings to the positive \emph{Q}-value
neutron transfer channels may play a role in the sub-barrier fusion
enhancement. A rather simple model calculation taking neutron transfers
into account is proposed to overcome the discrepancies of $^{32}$S+$^{96}$Zr.

\end{abstract}

\pacs{25.70.Gh, 25.70.Jj, 24.10.Eq}

\maketitle

\section{Introduction}

The heavy-ion fusion reactions in the low-energy range near and
below the Coulomb barrier have been the subject of extensive
experimental and theoretical efforts in the past
decades~\cite{Balantekin98, Dasgupta98,Canto06}. Beside the fact
that the questions of the possible occurence of unexpected
phenomena, such as breakup effects on the fusion reactions at near
barrier energies~\cite{Canto06}, are still unresolved, one has still
to understand better the role of neutron transfers in the fusion
process~\cite{Broglia83,Lee84}. For instance, effects of
neutron-rich projectiles on the formation of super-heavy elements
(SHE)~\cite{Zagrebaev07}, especially with the development of newly
available Radioactive Ion Beams (RIB) facilities need to be
clarified as well as fusion hindrance at extremely low energies that
remain among the most interesting open questions in the nuclear
astrophysics domain \cite{Smith01}. Fusion enhancement below the
Coulomb barrier is one of the most studied phenomena and,
measurements of fusion barrier distributions have been widely
performed to investigate the mutual importance of both the nuclear
structure and dynamical process effects on the sub-barrier fusion
enhancement
\cite{Rowley91,Wei91,Stefanini95,Timmers98,Trotta02,Stefanini06,Stefanini07}.

Coupled-channels (CC) calculations have been used to describe the
reactions in this energy range theoretically (see for example Refs.
\cite{Balantekin98,Rowley91} and references therein). Fusion enhancement due
to the static deformations and surface vibrations of the nuclei has been well
described in the coupled-channels calculations
\cite{Rowley91,Wei91,Stefanini95,Timmers98,Trotta02,Stefanini06,Stefanini07}.

The influence of the neutron transfer channels on sub-barrier fusion
process
\cite{Timmers98,Trotta02,Stefanini06,Stefanini07,Stelson88,Stelson90,Shapira93}
is not yet fully understood. During the last 20 years a large number
of experimental and theoretical investigations were undertaken to
study the neutron-transfer mechanisms in competition with the fusion
process. Stelson \emph{et al}.~\cite{Stelson88,Stelson90,Shapira93}
proposed an original scenario that uses an empirical method
involving a sequential transfer of several neutrons between the
reactants. This multineutron transfer process is capable to initiate
fusion at large internuclear distances and will smooth the fusion
barrier distributions (with larger width) with lower energy
thresholds. This ``shift" effect corresponds to the energy window
for which the nuclei are allowed to come sufficiently close together
for neutrons to flow freely between the target and projectile. As a
consequence, this will reduce the effective barrier and enhance the
fusion cross sections at sub-barrier energies. Following this idea,
Rowley \emph{et al}.~\cite{Rowley92} have first used a simple
phenomenological model that simulates coupling to neutron transfer
channels with a parametrized coupling matrix. Later on,
Zagrebaev~\cite{Zagrebaev03} proposed another semiclassical
theoretical model that has been successfully used to reproduce the
sub-barrier fusion enhancement of the old $^{40}$Ca+$^{96}$Zr
reaction data of Ref.~\cite{Timmers98} by including the intermediate
positive \emph{Q}-value neutron transfer channels in the CC
calculations.

The failure of the CC calculations including only the couplings to
the inelastic excitations indicates that couplings to neutron
transfer channels may play a key role in the fusion dynamics near
the barrier for medium-heavy systems such as $^{40}$Ca+$^{90,96}$Zr
\cite{Timmers98,Montagnoli02}, $^{36}$S+$^{90,96}$Zr
\cite{Stefanini00}, and $^{20}$Ne+$^{90,92}$Zr \cite{Piasecki09},
the last two reactions being studied by measurements of large-angle
quasielastic scattering. Previous measurement of quasielastic
scatterings of $^{32}$S+$^{90,96}$Zr were also undertaken at
backward angles near the barrier~\cite{Yang08}; their analysis gave
indication that positive \emph{Q}-value neutron transfer channels
should be included in the coupling scheme. Up to now no fusion data
exist neither for $^{20}$Ne+$^{90,92}$Zr nor for the
$^{32}$S+$^{90,96}$Zr reactions, it will be interesting to measure
their fusion excitation functions. In order to disentangle the
possible effect of positive \emph{Q}-value neutron transfer
couplings we decided to investigate the two last systems. We report
here about the measurement of near- and sub-barrier fusion
excitation functions of $^{32}$S+$^{90}$Zr and $^{32}$S+$^{96}$Zr
performed with small energy
steps and good statistics accuracy.\\

Our research will focus on the role of neutron transfers between the
colliding nuclei as a mechanism that enhances the fusion cross sections
at sub-barrier energies. This paper is organized as follows: Sec. II
presents the experimental setup and details on the measurements. Results
of the analysis of the experimental data are given in Sec. III. Their
discussion is finally proposed in Sec. IV in the framework of comparisons
with semiclassical coupled-channels calculations before a short summary
of the Sec. V.

\section{Experimental procedures}

The experiment was performed at the HI-13 tandem accelerator of
CIAE, Beijing. The collimated $^{32}$S (Q = 10$^{+}$ charge state)
beam was used to bombard the zirconium oxide targets. The beam
intensity was stabilized in the 2-20 pnA range in order to minimize
the pile-up for each of the bombarding energies. The 3 mm diameter
(98.87\% enriched) $^{90}$ZrO$_{2}$ and (86.4\% enriched)
$^{96}$ZrO$_{2}$ targets were both 50 $\muup$g/cm$^{2}$ thick and
evaporated onto 15 $\muup$g/cm$^{2}$ carbon foil backings. The beam
energies varied over the range $\emph{E}_{\mathrm{lab}}$ = 100-130
MeV for $^{32}$S+$^{90}$Zr and $\emph{E}_{\mathrm{lab}}$ = 95-130
MeV for $^{32}$S+$^{96}$Zr (in 1.33 Mev steps at the higher and 0.67
MeV at the lower energies) and changed only downwards starting at
$\emph{E}_{\mathrm{lab}}$ = 130 MeV in order to reduce the magnetic
hysteresis for both targets. The target chamber contains four
silicon detectors at $\theta$ = $\pm$25$\,^{\circ}$ symmetrically
(right/left and up/down) with respect to the beam axis in the
forward direction in order to monitor the beam optics (Rutherford
scattering) and to provide an absolute normalization of the fusion
cross sections.

The fused evaporation residues (ER) concentrated to within a few
degrees of the incident beam direction were separated from the
incident beam (see Fig. 1) by an electrostatic deflector which
design is pretty much similar to the experimental setup in
Legnaro~\cite{Beghini85}. It consists of two pairs of electrodes
followed by an \emph{E}-TOF arrangement with of a microchannel plate
(MCP) detector coupled to a Si(Au) surface barrier detector.
Two-dimensional plots of the data were used to cleanly separate the
ER's from the beam-like products (BLP). A typical example of the
time-of-flight versus energy spectrum for $^{32}$S+$^{96}$Zr
measured at $\emph{E}_{\mathrm{lab}}$ = 130 MeV and $\theta$ =
2$\,^{\circ}$ is shown in Fig. 1. The electrostatic deflector could
be rotated about the target position in the horizontal plane to
measure the ER angular distributions.

\begin{figure}
\includegraphics[width=3.0in,height=2.5in]{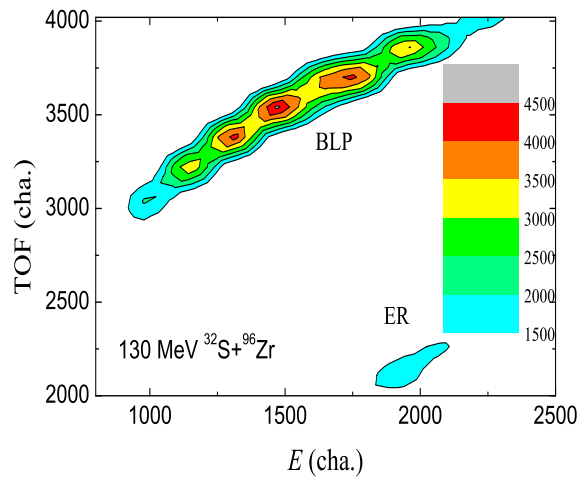}
\caption{\label{Spec} (Color online) Two-dimensional plot \emph{E}-TOF
of the events, following beam separation, taken at
$\emph{E}_{\mathrm{lab}}$ = 130 MeV and at 2$\,^{\circ}$ for the
$^{32}$S+$^{96}$Zr reaction. Two groups of particles [beam-like
particles(BLP) and evaporation residues (ER)] are indicated.}
\end{figure}

The particles coming from the target were selected before entering
the fields by an entrance collimator of 3 mm diameter, corresponding
to a $\Delta\theta$ = $\pm$ 0.57$\,^{\circ}$ opening. A 10$\mu$g/cm$^{2}$
thick carbon foil clung to the collimator was used to reset the atomic
charge state distribution on the ion path. The collimator of the MCP
defined the solid angle of the electrostatic deflector as being approximately
$\Delta\Omega$ = 0.3 msr.

ER angular distributions were measured in the range $\theta$ =
-4$\,^{\circ}$ to $\theta$ = 10$\,^{\circ}$ with step $\Delta\theta$
= 1$\,^{\circ}$ at three beam energies ($\emph{E}_{\mathrm{lab}}$ =
100, 115, and 130 MeV) for both systems. The angular distributions
were found to be symmetrical about $\theta$ = 0$\,^{\circ}$, as
expected. Their typical shape did not change appreciably with the
beam energy. These combined angular distributions and the double
Gaussian fits were used to obtain the fusion cross sections. At each
energy the number of ER events was normalized to the Rutherford
scattering rates counted by the monitor detectors. For the most of
energy points, only differential cross sections were measured at
$\theta$ = 2$\,^{\circ}$, from the obtained values, the total ER
cross sections were deduced. Using the solid angles, the  $\theta$ =
2$\,^{\circ}$-to-total ratios and the measured transmission
efficiencies, these ER yields were transformed into total cross
sections. Since fission of the compound nucleus can be neglected for
both systems, the measured cross sections were taken as complete
fusion cross section $\sigmaup$$_{f}$.

The transmission efficiencies and the relevant voltages used to
deflect the ER were calibrated by the $^{122}$Ba beam scattered by
the $^{90}$Zr target at small angles and at the corresponding
energies with the fusion evaporation residues. It was found that the
defocusing effect of the deflection voltage reduces the transmission
from unity to 0.60$\pm$0.06. Additional systematic errors come from
the geometrical solid angle uncertainties, the angular distribution
integrations, and the transmission measurements. Altogether these
contributions sum up to a $\pm$ 15\% value for systematic errors.

\section{Experimental results}

The measured ER excitation functions for the two systems are shown
in Fig. 2, where the energy scale is corrected for the target
thickness. The statistical errors shown in the figure do not exceed
the symbol size for most of the experimental points. They are $\pm$
0.8\% for both the high-energy and the intermediate-energy points
and increase to $\pm$ 23\% for the low-energy points. The ER cross
sections are listed in Tables I and II for both reactions.

\begin{table}
\caption{\label{}Experimental ER cross sections for
$^{32}$S+$^{90}$Zr.}

\begin{center}
\begin{tabular}{cccc}
\hline \hline

\emph{E}$_{\mathrm{c.m.}}$ (MeV) & $\sigma_{\mathrm{ER}}$ (mb) &
\emph{E}$_{\mathrm{c.m.}}$ (MeV) & $\sigma_{\mathrm{ER}}$ (mb)
\\
\hline

95.3 & 623.51$\pm$4.00 & 82.7 & 142.10$\pm$1.37\\
94.2 & 582.98$\pm$4.63 & 82.2 & 122.84$\pm$1.08\\
93.3 & 528.80$\pm$3.93 & 81.7 & 105.94$\pm$1.10\\
92.2 & 535.48$\pm$4.70 & 81.2 &  91.59$\pm$0.92\\
91.2 & 430.00$\pm$3.37 & 80.7 &  77.96$\pm$0.80\\
90.7 & 442.29$\pm$3.03 & 80.2 &  63.69$\pm$0.62\\
90.2 & 429.44$\pm$3.58 & 79.7 &  54.96$\pm$0.55\\
89.7 & 363.05$\pm$3.22 & 79.2 &  40.22$\pm$0.39\\
89.2 & 357.17$\pm$2.84 & 78.7 &  32.89$\pm$0.34\\
88.7 & 342.70$\pm$2.94 & 78.2 &  23.41$\pm$0.35\\
88.2 & 322.27$\pm$3.07 & 77.7 &  17.46$\pm$0.35\\
87.7 & 309.41$\pm$2.80 & 77.2 &  10.58$\pm$0.42\\
87.2 & 298.07$\pm$2.85 & 76.7 &   5.89$\pm$0.41\\
86.7 & 277.99$\pm$2.29 & 76.2 &   3.66$\pm$0.37\\
86.2 & 246.27$\pm$2.81 & 75.7 &   2.07$\pm$0.25\\
85.7 & 244.09$\pm$2.02 & 75.2 &   1.09$\pm$0.16\\
85.2 & 224.93$\pm$1.95 & 74.7 &   0.57$\pm$0.10\\
84.7 & 190.06$\pm$1.64 & 74.2 &   0.31$\pm$0.06\\
84.2 & 189.70$\pm$1.75 & 73.7 &   0.18$\pm$0.04\\
83.7 & 167.85$\pm$1.25 & 73.2 &   0.12$\pm$0.03\\
83.2 & 143.85$\pm$1.38 & \\
\hline
\hline

\end{tabular}
\end{center}
\vspace{-0.8cm}
\end{table}

\begin{table}
\caption{\label{}Experimental ER cross sections for
$^{32}$S+$^{96}$Zr.}

\begin{center}
\begin{tabular}{cccc}
\hline \hline

\emph{E}$_{\mathrm{c.m.}}$ (MeV) & $\sigma_{\mathrm{ER}}$ (mb) &
\emph{E}$_{\mathrm{c.m.}}$ (MeV) & $\sigma_{\mathrm{ER}}$ (mb)
\\
\hline

97.1 & 683.96$\pm$3.93 & 82.1 & 161.35$\pm$1.34\\
95.9 & 644.56$\pm$7.76 & 81.8 & 145.06$\pm$1.51\\
95.0 & 601.59$\pm$3.70 & 81.1 & 130.62$\pm$1.09 \\
93.9 & 591.22$\pm$4.33 & 80.8 & 117.57$\pm$1.21\\
93.0 & 505.50$\pm$3.00 & 80.1 &  99.28$\pm$0.85\\
92.3 & 519.80$\pm$4.10 & 79.7 &  90.53$\pm$0.83\\
92.0 & 510.25$\pm$3.57 & 79.1 &  72.13$\pm$0.66\\
91.3 & 439.19$\pm$3.38 & 78.7 &  66.85$\pm$0.70\\
90.9 & 448.78$\pm$3.14 & 78.1 &  50.41$\pm$0.43\\
90.3 & 381.53$\pm$2.95 & 77.7 &  46.28$\pm$0.51\\
89.9 & 424.93$\pm$3.13 & 77.0 &  37.01$\pm$0.31\\
89.3 & 382.96$\pm$3.02 & 76.7 &  31.81$\pm$0.34\\
88.9 & 393.91$\pm$2.55 & 76.0 &  23.82$\pm$0.03\\
88.3 & 354.74$\pm$2.11 & 75.7 &  20.18$\pm$0.03\\
88.0 & 338.00$\pm$2.74 & 75.0 &  15.32$\pm$0.23\\
87.2 & 314.40$\pm$2.44 & 74.6 &  13.12$\pm$0.26\\
86.9 & 313.05$\pm$2.37 & 74.0 &   8.95$\pm$0.36\\
86.2 & 264.38$\pm$2.37 & 73.6 &   5.53$\pm$0.39\\
85.8 & 282.11$\pm$2.51 & 73.0 &   2.95$\pm$0.30\\
85.2 & 244.76$\pm$1.78 & 72.5 &   1.94$\pm$0.23\\
84.8 & 231.95$\pm$2.18 & 71.9 &   0.79$\pm$0.12\\
84.2 & 222.32$\pm$1.83 & 71.4 &   0.69$\pm$0.12\\
83.8 & 231.63$\pm$1.93 & 70.9 &   0.37$\pm$0.07\\
83.2 & 182.87$\pm$1.85 & 70.4 &   0.23$\pm$0.05\\
82.8 & 179.30$\pm$1.82 & 69.9 &   0.09$\pm$0.02\\
\hline
\hline

\end{tabular}
\end{center}
\vspace{-0.8cm}
\end{table}

\begin{figure}
\includegraphics[width=3.0in,height=2.5in]{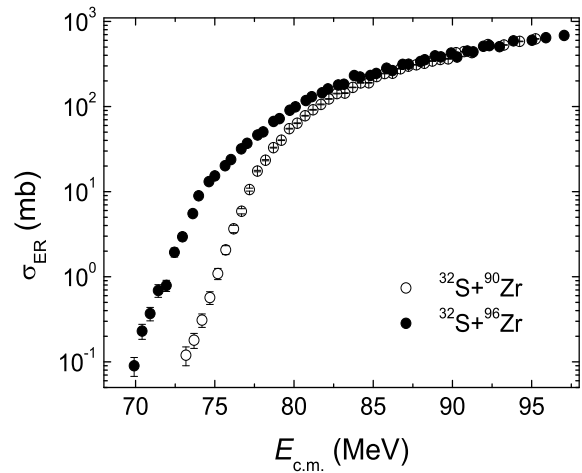}
\caption{\label{Fusi} Experimental ER excitation functions of
$^{32}$S+$^{90}$Zr (hollow circles) and $^{32}$S+$^{96}$Zr (solid
circles) as a function of the center-of-mass energy. The error bars
represent purely statistical uncertainties. }
\end{figure}

\begin{figure}
\includegraphics[width=3.0in,height=2.5in]{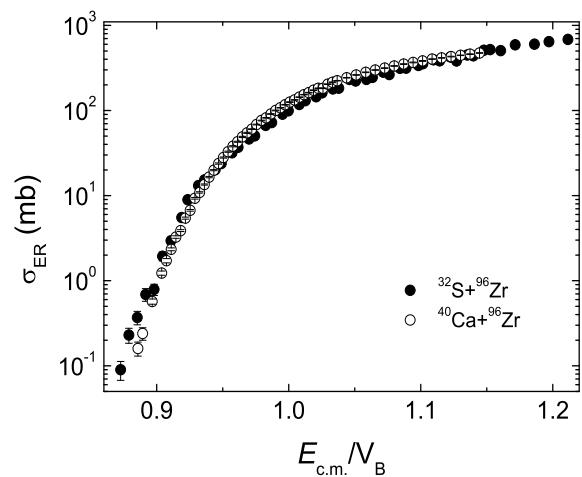}
\caption{\label{Comp} Plot of the ER excitation functions on a
reduced-energy scale for the two systems $^{32}$S+$^{96}$Zr (filled
circles) measured in the present work and $^{40}$Ca+$^{96}$Zr (open
circles). The data for $^{40}$Ca+$^{96}$Zr are taken from
Ref.~\cite{Timmers98}.}
\end{figure}

The comparison of the ER excitation functions of $^{32}$S+$^{96}$Zr
(present work) and $^{40}$Ca+$^{96}$Zr ~\cite{Timmers98} is made
easier when cross sections are plotted in a reduced-energy scale as
shown in Fig. 3. One observes that the two systems display very
similar behaviors on the whole energy range despite relatively large
discrepancies. These discrepancies are mainly due to larger
uncertainties in the present data arising from larger backgrounds in
the spectra. It is interesting to notice that both system reactants
have very similar nuclear structures as well as neutron-transfer
properties. This behavior, already discussed in our previous
investigation of $^{32}$S+$^{90,96}$Zr quasielastic barrier
distributions \cite{Yang08}, indicates that the positive
$\emph{Q}$-value neutron transfers strongly enhance the fusion cross
sections at sub-barrier energies. This experimental observation will
be confirmed by the semiclassical coupled-channels calculations as
discussed in the following Section.

\section{Discussion: semiclassical coupled-channels calculations}

The ER excitation functions of $^{32}$S+$^{90}$Zr and
$^{32}$S+$^{96}$Zr have been compared with the results of CC
calculations performed with the CCFULL code \cite{CCFULL} using
the Aky\"{u}z-Winther nuclear potential \cite{AW}.

The relevant informations on the low-lying excitations of
$^{32}$S, $^{90}$Zr, and $^{96}$Zr can be seen in Table III. The
quadrupole vibrations of both $^{90}$Zr and $^{96}$Zr nuclei are
weak in energy: in fact, they lie at comparable energies. With
this potential the CCFULL barriers were found to be at V$_B$ =
81.2 MeV for $^{32}$S+$^{90}$Zr and at V$_B$ = 80.1 MeV for
$^{32}$S+$^{96}$Zr, respectively. These values are fully
consistent with what was found for the quasielastic barriers
previously measured by our group \cite{Yang08}. In a first
approximation, these values can be considered as average values
between the barrier heights for the nose-to-nose configuration and
for the side-to-side configuration~\cite{Zagrebaev04}.

Figures 4 and 5 show the comparisons of the experimental ER
excitation functions and the CCFULL calculations with one- and
two-phonon couplings and without coupling for the
$^{32}$S+$^{90,96}$Zr fusion reactions. Semiclassical
coupled-channels calculations have also been performed by use of
the new code NTFus \cite{Axel} where the proximity potential is
adopted. The new oriented object code has been constructed in the
framework of the Zagrebaev model~\cite{Zagrebaev04} and
implemented in C$^{++}$ by using the compiler of ROOT~\cite{Root}.
More technical details and related discussions about the NTFus
code will be illustrated elsewhere \cite{Axel}.

\begin{figure}
\includegraphics[width=3.2in,height=2.3in]{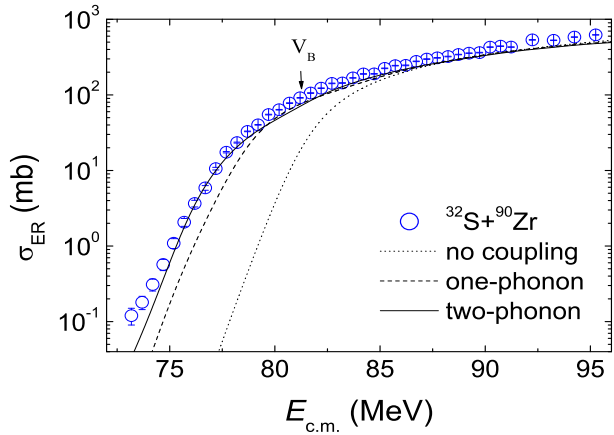}
\caption{\label{Calc} ER excitation function of
$^{32}$S+$^{90}$Zr. The open circles are the experimental data.
The dotted, dashed, and solid lines represent the CCFULL
calculations without coupling and with the one-phonon and
two-phonon couplings, respectively (see text for details). The
arrow indicates the position of the Coulomb barrier for
$^{32}$S+$^{90}$Zr.}
\end{figure}

\begin{figure}
\includegraphics[width=3.2in,height=2.3in]{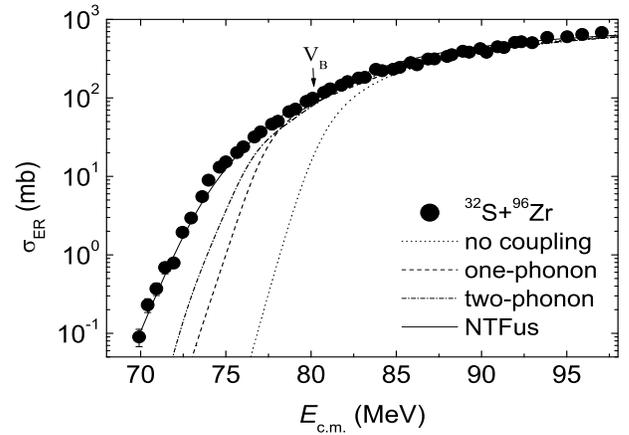}
\caption{\label{Calc} ER excitation functions of $^{32}$S+$^{96}$Zr.
The solid circles are the experimental data. The dotted, dashed and
dash-point lines are the CCFULL calculations with no coupling,
one-phonon and two-phonon couplings, respectively. The solid line is
the calculation taking into account the neutron transfers by NTFus
code. The arrow indicates the position of the Coulomb barrier for
$^{32}$S+$^{96}$Zr.}
\end{figure}

First we propose to present the calculations for
$^{32}$S+$^{90}$Zr that are displayed in Fig. 4. They were
performed without taking the neutron transfers into account. The
two-phonon coupling CCFULL calculations are quite satisfactory.
The calculation (solid line) reproduces the data below and above
the barrier V$_B$ (arrow in Fig.4). On the other hand, the
calculations fail for $^{32}$S+$^{96}$Zr with large discrepancies
occuring mainly at energies below the barrier V$_B$ (arrow in
Fig.5). One- and two-phonon (both shown in Fig. 5) excitations in
$^{96}$Zr, of both quadrupole and octupole natures, bring
additional but not sufficient enhancements. Finally, we tried also
"three-phonon" couplings, but no further improvement could be
reached. Anyway, additional couplings in $^{32}$S+$^{96}$Zr, which
might give rise to lower-energy barriers, are simply not present
in the coupling scheme. Similar conclusions have been obtained for
the $^{40}$Ca+$^{94}$Zr reaction \cite{Stefanini07}.

When we choose to take into account the neutron transfers, the
fusion excitation function can be derived using the following
formula~\cite{Zagrebaev03}:

\setcounter{equation}{0}
\begin{eqnarray}
T_{l}(E)&=& \int f(B)\frac{1}{N_{\mathrm{tr}}} \sum_{k}
\int_{-E}^{Q_0(k)}\alpha_k(E,l,Q) \nonumber\\
&& \times P_{HW}(B,E+Q,l)dQdB,
\end{eqnarray}
and
\begin{eqnarray}
\sigma_{fus}(E)= \frac{\pi\hbar^{2}}{2\mu E}
\sum\limits_{l=0}^{l_{cr}}(2l+1)T_l(E),
\end{eqnarray}
where \emph{T}$_{l}$(\emph{E}) is the transmission, \emph{E} the
energy at the center-of-mass, \emph{f}(\emph{B}) the normalized
barrier distribution function, \emph{l} the momentum and
\emph{l}$_{\mathrm{cr}}$ the critical momentum calculated where
there is no coupling (well above the barrier).
$\alpha_{\mathrm{k}}$(\emph{E},\emph{l},\emph{Q}) and
$\emph{Q}_{\mathrm{0}}$(k) are the probability and the
\emph{Q}-value for the transfer of k neutrons, 1/N$_{tr}$ is the
normalization of the total probability taking into account the
neutron transfers.

The calculation with the neutron transfer effect is performed up to
the channel +4n (k=4). No more visible effect can be obtained by
using +5n and +6n channels. More details of the calculation
procedures and of the description of the NTFus code~\cite{Axel},
itself, will be given in a forthcoming Brief Report. The
\emph{Q}-values for the calculation (solid line in Fig. 5) are given
in Table IV. As we can see in Fig. 5, the dash-point line (without
the neutron transfers) does not at all describe the data at the
sub-barrier energies. In contrast, the solid line taking into
account the neutron transfers is able to fit the data reasonably
well. As expected, the correction applied on the calculation at
sub-barrier energies by the Zagrebaev model
\cite{Zagrebaev03,Zagrebaev04} enhances the cross sections further.
Moreover, it allows a fairly good description of the present
experimental data showing the strong effect of neutron transfers for
the sub-barrier fusion of $^{32}$S+$^{96}$Zr.

\begin{table}
\caption{\label{tab1} Excitation energies $\emph{E}_{\mathrm{x}}$,
spin and parities $\lambda^{\pi}$, and deformation parameters
$\beta_{\lambda}$ for $^{32}$S and $^{90,96}$Zr.}
\begin{tabular}{cccc}
\hline
\hline
Nucleus & $\emph{E}_{\mathrm{x}}$ (MeV) & $\lambda^{\pi}$ & $\beta_{\lambda}$ \\
 \hline
$^{32}$S  &  2.230 &  2$^{+}$ &  0.32 \\
          &  5.006 &  3$^{-}$ &  0.40 \\
$^{90}$Zr &  2.186 &  2$^{+}$ &  0.09 \\
          &  2.748 &  3$^{-}$ &  0.22 \\
$^{96}$Zr &  1.751 &  2$^{+}$ &  0.08 \\
          &  1.897 &  3$^{-}$ &  0.27 \\

\hline
\hline
\end{tabular}
\end{table}

\begin{table}
\caption{\label{tab2} \emph{Q}-value in MeV for neutron pickup
transfer channels from ground state to ground state for the
$^{32}$S+$^{90,96}$Zr systems.}
\begin{center}
\begin{tabular}{ccccc}
\hline
\hline
   System &  +1n  &  +2n &  +3n  &  +4n \\
\hline
$^{32}$S+$^{90}$Zr &  -3.33 &  -1.229 &  -6.59) &  -6.319\\
$^{32}$S+$^{96}$Zr &  0.788 &  5.737  &  4.508  &  7.655\\
\hline
\hline
\end{tabular}
\end{center}
\end{table}

  Figures 6 and 7 show the experimental barrier distributions from
fusion and quasi-elastic scattering and the corresponding CCFULL
calculations for the two systems. The fusion barrier distributions
for the two systems have been obtained by double differentiation
\emph{E}$\sigma_{\mathrm{fus}}$ vs energy using the three-point
difference formula \cite{Timmers98}. The quasi-elastic barrier
distributions are taken from Ref. \cite{Yang08}. It is very
interesting to note that for both reactions the experimental
quasi-elastic barrier distributions and the experimental fusion
barrier distributions are strikingly similar. The large
fluctuations occurs in the barrier distributions for
\emph{E}$_{\mathrm{c.m.}}>$80 MeV is not enough due to the
measuring accuracy of the ER cross sections. For
$^{32}$S+$^{90}$Zr, the overall trends of the experimental barrier
distributions are roughly consistent with the CCFULL calculation
considering the two-phonon coupling. While for $^{32}$S+$^{96}$Zr,
the experimental barrier distributions are wide and show a
low-energy tail extending to the lowest energies compared with
$^{32}$S+$^{90}$Zr and the CCFULL calculation considering the
two-phonon coupling. It shows a part loss of the component below
75 MeV. This is due to the coupling to the $\emph{Q}>$0 neutron
transfers, corresponding to the further fusion enhancement at
sub-barrier energies compared with the calculation.

\begin{figure}
\includegraphics[width=3.2in,height=2.3in]{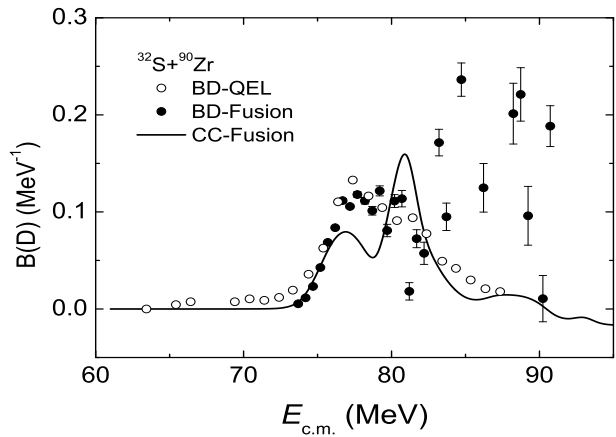}
\caption{\label{Calc} Barrier distributions for $^{32}$S+$^{90}$Zr
from fusion (solid circles) and quasi-elastic scattering (hollow
circles). The line is the CCFULL calculation with two-phonon
coupling.}
\end{figure}

\begin{figure}
\includegraphics[width=3.2in,height=2.3in]{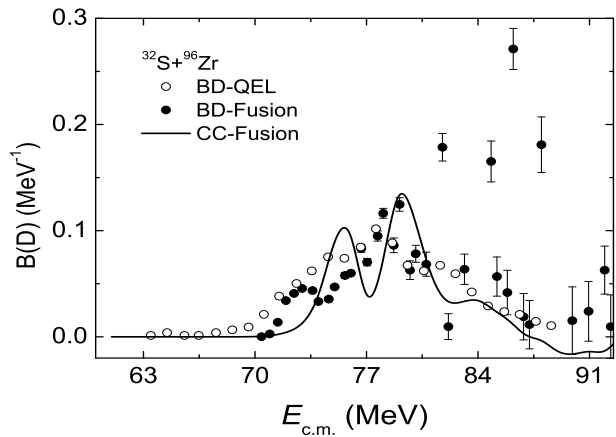}
\caption{\label{Calc} For $^{32}$S+$^{96}$Zr, the symbols are the
same as in Fig. 6.}
\end{figure}

\section{Summary}

The fusion excitation functions for $^{32}$S+$^{90,96}$Zr were
measured with a high precision near and below the Coulomb barrier.
The sub-barrier cross sections for $^{32}$S+$^{96}$Zr are much
larger compared with $^{32}$S+$^{90}$Zr. The data have been
analyzed with both the CCFULL code and the NTFus code based on a
semi-classical coupled-channels model, which includes the
sequential neutron transfers for $^{32}$S+$^{96}$Zr as earlier
proposed by Zagrebaev. Good agreement between experimental data
and the calculation is achieved for $^{32}$S+$^{90}$Zr by
including the couplings to the low-lying quadruple and octupole
vibrations in $^{32}$S and $^{90}$Zr. Whereas the NTFus
calculations can reproduce the data by including four sequential
neutron transfer channels as well as the low-lying quadrupole and
octupole vibrations in $^{32}$S and $^{96}$Zr. The comparison with
previous data on $^{40}$Ca+$^{96}$Zr shows that the excitation
functions of $^{32}$S, $^{40}$Ca+$^{96}$Zr are very similar in the
whole energy range. Both systems have similar collective states
and positive \emph{Q}-value neutron transfer channels. This
comparison strongly supports the previous suggestion \cite{Yang08}
that positive \emph{Q}-value neutron transfer channels enhance
sub-barrier fusion cross sections, particularly at very low
energies. Also the fusion and quasi-elastic barrier distributions
of the $^{32}$S+$^{90,96}$Zr systems are basically consistent in
both cases. The two barrier distributions for $^{32}$S+$^{96}$Zr
cannot be reproduced by CCFULL code below 75 MeV. The fact shows
again the effect of the $\emph{Q}>$0 neutron transfers on the
sub-barrier fusion process. In addition to the fusion excitation
function, the neutron transfer cross section measurement for this
system should provide useful information on the coupling strength
of neutron transfer channels, which will allow us to reach a much
deeper understanding of the role of neutron transfer mechanisms,
sequential or simultaneous, in the fusion process.

\begin{acknowledgments}

This work was supported by the National Natural Science Foundation of
China under Grant Nos. 10575134, 10675169, 10735100, and the Major
State Basic Research Developing Program under Grant No. 2007CB815003
as well as the Boussole Grant 2009 No. 080105519 from Region Alsace
(France) received by A.R. Two of us (A.R. and C.B.) would like to
thank all the members of the experimental team of CIAE Beijing for
their very kind hospitality and assistance.

\end{acknowledgments}

\end{document}